\documentclass{article}
\usepackage[preprint]{neurips_2026}         %

\usepackage[utf8]{inputenc}
\usepackage[T1]{fontenc}
\usepackage{hyperref}
\usepackage{url}
\usepackage{booktabs}
\usepackage{amsfonts}
\usepackage{amsmath}
\usepackage{microtype}
\usepackage{xcolor}
\usepackage{graphicx}
\usepackage{multirow}
\usepackage{makecell}
\usepackage{array}
\usepackage[capitalize,noabbrev]{cleveref}
\usepackage{colortbl}
\usepackage{tcolorbox}
\tcbuselibrary{breakable}
\usepackage{listings}
\usepackage{xspace}
\usepackage{tikz}
\usetikzlibrary{arrows.meta, positioning, backgrounds, fit, calc}
\usepackage{enumitem}
\usepackage{wrapfig}
\usepackage{pgfplots}
\pgfplotsset{compat=1.18}

\definecolor{lightgray}{RGB}{248,249,250}
\definecolor{color-b}{RGB}{129, 178, 154}
\definecolor{color-d}{RGB}{242, 204, 143}
\definecolor{color-e}{RGB}{224, 122, 95}
\definecolor{light-gray}{gray}{0.4}

\newcommand{\bench}{\textsc{TraceEval}\xspace}
\newcommand{\code}[1]{\texttt{\small #1}}

\newcolumntype{L}[1]{>{\raggedright\let\newline\\\arraybackslash\hspace{0pt}}m{#1}}
\newcolumntype{C}[1]{>{\centering\let\newline\\\arraybackslash\hspace{0pt}}m{#1}}
\newcolumntype{R}[1]{>{\raggedleft\let\newline\\\arraybackslash\hspace{0pt}}m{#1}}

\lstdefinestyle{codestyle}{
    basicstyle=\ttfamily\footnotesize,
    breaklines=true,
    frame=single,
    numbers=left,
    numberstyle=\tiny\color{gray},
    backgroundcolor=\color{lightgray},
    xleftmargin=1.5em,
    framexleftmargin=1em,
}

\title{An Execution-Verified Multi-Language Benchmark \\for Code Semantic Reasoning}

\author{%
  \textbf{Yikun Li$^{1}$ \quad Jinfeng Jiang$^{1}$ \quad Ting Zhang$^{2}$ \quad Chengran Yang$^{1}$ \quad Chenxing Zhong$^{3}$} \\
  \textbf{Yin Yide$^{4}$ \quad Leow Wen Bin$^{4}$ \quad Eng Lieh Ouh$^{1}$ \quad Lwin Khin Shar$^{1}$ \quad David Lo$^{1}$} \\[0.5em]
  \normalfont%
  $^{1}$Singapore Management University \quad $^{2}$Monash University \\
  $^{3}$Nanjing University of Science and Technology \quad $^{4}$GovTech Singapore
}

\begin{document}
\maketitle

\begin{abstract}
Evaluating whether large language models (LLMs) can recover execution-relevant program structure, rather than only produce code that passes tests, remains an open problem.
Existing code benchmarks emphasize test-passing outputs, from standalone programming tasks (HumanEval, MBPP, LiveCodeBench) to repository repair (SWE-Bench); this is useful, but offers limited diagnostic signal about which program semantics a model can recover from source.
We introduce \bench, to our knowledge the first execution-verified, multi-language benchmark for \emph{code semantic reasoning}: recovering a program's runtime call structure from source code.
Unlike prior call-graph benchmarks that rely on static-tool output or hand-annotated ground truth, every positive edge in \bench is \emph{mechanically witnessed} by validation execution, eliminating annotator disagreement and label noise for observed behavior.
\bench consists of (i) 10{,}583 real-world programs (2{,}129 test, 8{,}454 train) extracted from 1{,}600+ open-source repositories across Python, JavaScript, and Java via an LLM-assisted harness-generation pipeline with tracer validation; and (ii) a reproducible pipeline that converts any open-source repository into new verified benchmark instances.
We evaluate 10 LLMs at zero-shot on the held-out test split.
The strongest model, Claude-Opus-4.6, reaches an average F1 of $\mathbf{72.9\%}$ across the three languages.
To demonstrate the train split's utility as a supervision substrate, we fine-tune the Qwen2.5-Coder family on it: lifts of up to $\mathbf{+55.6}$ F1 bring tuned Qwen2.5-Coder-32B to $\mathbf{71.2\%}$, within $\mathbf{1.7}$ F1 of zero-shot Claude-Opus-4.6.
We release the benchmark, pipeline, baselines, and a datasheet at \url{https://github.com/yikun-li/TraceEva}.
\end{abstract}

\section{Introduction}
\label{sec:introduction}

The code-LLM evaluation landscape is dominated by test-passing tasks: standalone code generation in HumanEval~\citep{chen2021codex}, MBPP~\citep{austin2021mbpp}, and LiveCodeBench~\citep{jain2024livecodebench}, and repository-level issue resolution in the SWE-Bench family~\citep{jimenez2024swebench, openai2024swebenchverified, deng2025swebenchpro}.
These tasks measure whether a model can produce a solution or patch that satisfies tests, a useful capability, but one that conflates semantic reasoning with API knowledge, pattern retrieval, and test-specific search.
They also offer limited visibility into which execution-relevant program structures a model can recover from source.
Frontier LLMs now exceed $90\%$ on HumanEval, yet can still hallucinate call targets, misresolve aliases, and confuse virtual dispatch on straightforward programs.
For ML researchers studying reasoning, this is the same compositional-generalization question that motivates math and planning benchmarks, except code admits a mechanical, deterministic oracle for executed behavior: the runtime itself.
The resulting gap between test-passing performance and execution-grounded structure recovery remains undermeasured.

\paragraph{Call Graphs as a Probe of Semantic Reasoning.}
We argue that \emph{call graph construction} (predicting which functions call which, given source code) is a compact and diagnostic probe of code semantic reasoning.
It has three properties that make it well suited for a benchmark:
(i)~\textit{Multi-hop compositional reasoning.} Each predicted edge chains variable tracking, alias resolution, and dynamic dispatch: the same compositional reasoning skills that drive progress on math and planning benchmarks.
(ii)~\textit{Mechanically verifiable ground truth.} Under a fixed validation input set, positive edges are not subject to rater disagreement or LLM-as-judge bias: they are \emph{observed} during program execution.
(iii)~\textit{Downstream importance.} Call graph understanding supports common code-agent workflows such as navigation, refactoring, impact analysis, and targeted test selection.

Despite these properties, existing call-graph benchmarks do not bring them together.
Prior call graph benchmarks are each language-siloed (PyCG~\citep{salis2021pycg} for Python, CATS~\citep{reif2016cats} for Java, the JavaScript suite of \citet{venkatesh2025llm} for JavaScript) and rely on hand-annotated ground truth that can contain errors; we identified errors in PyCG while developing our tracer infrastructure.
Their corpora are also too narrow to support modern fine-tuning protocols.

\bench is positioned along two axes that distinguish it from existing code benchmarks (\Cref{tab:benchmark_comparison}): \emph{what capability is probed} (test-passing output vs.\ semantic reasoning) and \emph{how ground truth is established} (judgment vs.\ execution).
On the first axis, benchmarks such as HumanEval, MBPP, LiveCodeBench, and the SWE-Bench family score whether generated code or patches pass tests; \bench instead requires the model to recover the execution-covered caller$\rightarrow$callee relation from source.
On the second axis, prior benchmarks rely on four oracle types: test cases (which overfit), human rubrics like MT-Bench~\citep{zheng2023llmjudge} (which drift), LLM-as-judge protocols (biased), and static-tool ground truth from analyzers such as Joern~\citep{joern} and Jelly~\citep{jelly} (which trade recall for precision and disagree across tools~\citep{sui2020recall, antal2023jscg}).
\bench uses a fifth, orthogonal oracle, \emph{direct runtime observation}, so labels are reproducible, auditable, and free of rater variance.
No prior work combines a semantic reasoning task, execution-verified labels, multi-language coverage, and a reproducible pipeline for growing the benchmark.

\begin{table}[t]
\centering
\small
\caption{Positioning of \bench against representative code benchmarks. Size is the count of benchmark instances at release; for growable benchmarks the number reflects the snapshot reported in the cited paper. ``Semi'' denotes curated follow-on releases rather than a released benchmark-generation pipeline.}
\label{tab:benchmark_comparison}
\resizebox{\linewidth}{!}{%
\begin{tabular}{lcccrc}
\toprule
\textbf{Benchmark} & \textbf{Task Type} & \textbf{Ground Truth} & \textbf{Languages} & \textbf{Size} & \textbf{Growable} \\
\midrule
HumanEval~\citep{chen2021codex}      & Code Gen.      & Test Cases     & Python       & $164$        & No \\
MBPP~\citep{austin2021mbpp}          & Code Gen.      & Test Cases     & Python       & $974$        & No \\
\midrule
SWE-Bench~\citep{jimenez2024swebench} & Repo Repair   & Test Cases     & Python       & $2{,}294$    & Semi \\
SWE-Bench Pro~\citep{deng2025swebenchpro} & Repo Repair & Test Cases     & Python       & $1{,}865$    & Semi \\
SWE-Bench Verified~\citep{openai2024swebenchverified} & Repo Repair & Test Cases & Python  & $500$        & No \\
\midrule
LiveCodeBench~\citep{jain2024livecodebench} & Problem Solving & Test Cases & Python  & $511$        & Yes \\
\midrule
PyCG~\citep{salis2021pycg} & Call Graph & Hand-Annotated & Python  & $112$        & No \\
CATS~\citep{reif2016cats}            & Call Graph     & Hand-Annotated & Java         & $128$    & No \\
\midrule
\textbf{\bench (Ours)}               & Call Graph     & \textbf{Execution} & \textbf{Py / JS / Java} & $\mathbf{10{,}583}$ & \textbf{Yes} \\
\bottomrule
\end{tabular}%
}
\end{table}

\begin{figure*}[t]
\centering
\resizebox{0.96\textwidth}{!}{
\begin{tikzpicture}[
    box/.style={rectangle, draw=#1!60, fill=#1!10, rounded corners=2pt,
        minimum height=0.7cm, minimum width=2.2cm, align=center,
        font=\footnotesize\sffamily, line width=0.6pt},
    box/.default=black,
    plannedbox/.style={rectangle, draw=#1!50, fill=#1!4, rounded corners=2pt,
        minimum height=0.7cm, minimum width=2.2cm, align=center,
        font=\footnotesize\sffamily, line width=0.5pt, dashed},
    plannedbox/.default=black,
    stagebox/.style={rectangle, draw=#1!40, fill=#1!3, rounded corners=5pt,
        inner sep=7pt, line width=1pt},
    stagebox/.default=black,
    plannedstagebox/.style={rectangle, draw=#1!35, fill=#1!2, rounded corners=5pt,
        inner sep=7pt, line width=0.8pt, dashed},
    plannedstagebox/.default=black,
    arr/.style={-{Stealth[length=2.5mm, width=1.8mm]}, line width=0.8pt, color=#1!70},
    arr/.default=black,
    bigarr/.style={-{Stealth[length=3mm, width=2.2mm]}, line width=1.4pt, color=#1!80},
    bigarr/.default=black,
    plannedarr/.style={-{Stealth[length=3mm, width=2.2mm]}, line width=1pt, color=#1!50, dashed},
    plannedarr/.default=black,
    lbl/.style={font=\footnotesize\sffamily\bfseries, color=#1!75!black},
    lbl/.default=black,
    plannedlbl/.style={font=\footnotesize\sffamily\bfseries\itshape, color=#1!60},
    plannedlbl/.default=black,
    annot/.style={font=\tiny\sffamily, color=black!50, align=center},
]

\definecolor{cA}{RGB}{129, 178, 154}  %
\definecolor{cB}{RGB}{61, 64, 91}     %
\definecolor{cC}{RGB}{224, 122, 95}   %
\definecolor{cE}{RGB}{70, 120, 200}   %
\definecolor{cF}{RGB}{242, 204, 143}  %

\node[lbl=cA] (lbl1) at (0, 0) {Stage 1: Harness Generation};
\node[box=cA, below=0.20cm of lbl1, minimum width=4.6cm] (repos) {1{,}600+ Permissive GitHub Repos};
\node[box=cA, below=0.18cm of repos, minimum width=4.6cm] (rew) {LLM-Assisted Driver + Stubs};
\node[box=cA, below=0.18cm of rew, minimum width=4.6cm, line width=0.9pt, font=\footnotesize\sffamily\bfseries] (prog) {Executable Harnesses};
\draw[arr=cA] (repos) -- (rew);
\draw[arr=cA] (rew) -- (prog);
\begin{scope}[on background layer]
\node[stagebox=cA, fit=(lbl1)(repos)(rew)(prog), inner ysep=8pt] (s1) {};
\end{scope}

\node[lbl=cB] (lbl2) at (6.5, 0) {Stage 2: Dynamic Tracing};
\node[box=cB, below=0.20cm of lbl2, minimum width=4.6cm] (mktracer) {Tracers for Python / JS / Java};
\node[box=cB, below=0.18cm of mktracer, minimum width=4.6cm] (runtracer) {Run Tracers on Programs};
\node[box=cB, below=0.18cm of runtracer, minimum width=4.6cm, line width=0.9pt, font=\footnotesize\sffamily\bfseries] (edges) {Observed Caller$\rightarrow$Callee Edges};
\draw[arr=cB] (mktracer) -- (runtracer);
\draw[arr=cB] (runtracer) -- (edges);
\begin{scope}[on background layer]
\node[stagebox=cB, fit=(lbl2)(mktracer)(runtracer)(edges), inner ysep=8pt] (s2) {};
\end{scope}

\node[lbl=cC] (lbl3) at (13.0, 0) {Stage 3: Validation};
\node[box=cC, below=0.20cm of lbl3, minimum width=4.6cm] (exec) {Executable Under Driver};
\node[box=cC, below=0.18cm of exec, minimum width=4.6cm] (accept) {Accept Iff $\geq 2$ Traced Edges};
\node[box=cC, below=0.18cm of accept, minimum width=4.6cm, line width=0.9pt, font=\footnotesize\sffamily\bfseries] (corpus) {\bench: 10{,}583 Programs};
\draw[arr=cC] (exec) -- (accept);
\draw[arr=cC] (accept) -- (corpus);
\begin{scope}[on background layer]
\node[stagebox=cC, fit=(lbl3)(exec)(accept)(corpus), inner ysep=8pt] (s3) {};
\end{scope}

\draw[bigarr=cA] (s1.east) -- node[annot, above, yshift=1pt] {Programs} (s2.west);
\draw[bigarr=cB] (s2.east) -- node[annot, above, yshift=1pt] {Traces} (s3.west);

\node[box=cE, minimum width=4.6cm] (llm) at (22.0, 0.55) {10 LLMs (5 Frontier, 5 Open-Weight)};
\node[lbl=cE, above=0.20cm of llm] (lblE) {Zero-Shot Evaluation};
\node[box=cE, below=0.18cm of llm, minimum width=4.6cm, line width=0.9pt, font=\footnotesize\sffamily\bfseries] (score) {Edge-Level Scoring};
\draw[arr=cE] (llm) -- (score);
\begin{scope}[on background layer]
\node[stagebox=cE, fit=(lblE)(llm)(score), inner xsep=7pt, inner ysep=6pt, minimum width=5.4cm] (sE) {};
\end{scope}

\node[box=cF, minimum width=4.6cm] (base) at (22.0, -2.4) {Qwen2.5-Coder (1.5B / 7B / 32B)};
\node[lbl=cF, above=0.20cm of base] (lblF) {Supervised Fine-Tuning};
\node[box=cF, below=0.18cm of base, minimum width=4.6cm] (rt) {Reasoning Trace Generation};
\node[box=cF, below=0.18cm of rt, minimum width=4.6cm, line width=0.9pt, font=\footnotesize\sffamily\bfseries] (ft) {Tuned on Train Split};
\draw[arr=cF] (base) -- (rt);
\draw[arr=cF] (rt) -- (ft);
\begin{scope}[on background layer]
\node[stagebox=cF, fit=(lblF)(base)(rt)(ft), inner xsep=7pt, inner ysep=6pt, minimum width=5.4cm] (sF) {};
\end{scope}

\draw[bigarr=cC] (s3.east) to[out=18,in=180] node[annot, above, sloped, pos=0.48] {Test} (sE.west);
\draw[bigarr=cC] (s3.east) to[out=-18,in=180] node[annot, below, sloped, pos=0.50] {Train} (sF.west);

\end{tikzpicture}
}
\caption{Overview of the \bench pipeline. \textbf{Construction} (Stages 1--3): an LLM-assisted harness generator adds drivers and lightweight dependency stubs to files from $1{,}600+$ GitHub repositories, making them self-contained and executable; language-specific dynamic tracers execute those programs and observe caller$\rightarrow$callee edges; validation accepts only programs that execute successfully and produce at least two traced edges, yielding the $10{,}583$-program corpus ($2{,}129$ test / $8{,}454$ train). \textbf{Use}: the test split feeds zero-shot evaluation and failure analysis of ten LLMs, while the train split feeds LoRA fine-tuning of the Qwen2.5-Coder family.}
\label{fig:overview}
\end{figure*}
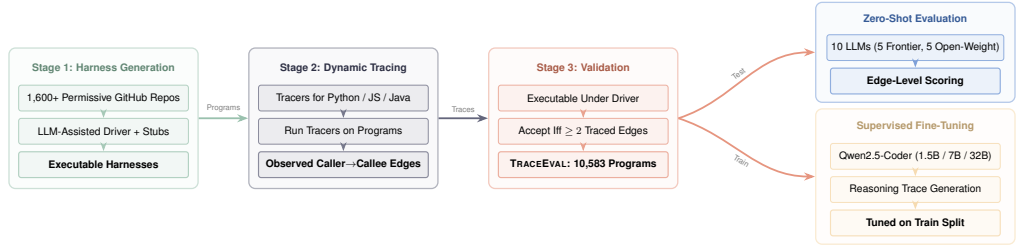

\paragraph{\bench: Execution as the Oracle.}
We present \bench, to our knowledge the first execution-verified call graph benchmark spanning Python, JavaScript, and Java.
The pipeline (\Cref{fig:overview}) pairs language-specific dynamic tracers (Python \code{sys.settrace}, JavaScript AST instrumentation, Java source-level method injection) with an LLM-assisted harness generator that adds drivers and lightweight dependency stubs to real-world source files from $1{,}600+$ permissively-licensed GitHub repositories, yielding a corpus of $10{,}583$ executable programs at roughly two orders of magnitude greater scale than prior call-graph benchmarks.
Every positive edge label is mechanically witnessed by validation execution; \bench therefore defines ground truth as the execution-covered call graph.

\paragraph{Overall Results.}
The strongest model, Claude-Opus-4.6, reaches an average F1 of only $\mathbf{72.9\%}$ across the three languages, while the weakest, Qwen2.5-Coder-1.5B, reaches $\mathbf{9.5\%}$, a $\mathbf{7.7{\times}}$ spread.
A failure-mode audit of the strongest model isolates three distinct mechanisms behind its residual error: untaken-branch hallucination, declared- versus runtime-type dispatch confusion, and class-name-as-callee output-schema mismatch.
Supervised LoRA fine-tuning on the released training split lifts the Qwen2.5-Coder family by up to $\mathbf{+55.6}$ F1, with tuned Qwen2.5-Coder-32B reaching $\mathbf{71.2\%}$, within $\mathbf{1.7}$ F1 of zero-shot Claude-Opus-4.6.
Together, these results show that the benchmark is diagnostic, exposing concrete reasoning failures that aggregate test-pass rates would hide.

\paragraph{Key Contributions.}
Our work makes four primary contributions:

\begin{itemize}[leftmargin=*, itemsep=2pt]
    \item \textbf{We introduce \bench}, a multi-language execution-verified call-graph corpus of $10{,}583$ real-world programs ($3{,}820$ Py, $3{,}226$ JS, $3{,}537$ Java) drawn from $1{,}600+$ open-source repositories under a unified JSON schema, partitioned into a $2{,}129$-program held-out test split and an $8{,}454$-program training split with no source-repository overlap (\Cref{sec:realworld}).
    \item \textbf{We release a reproducible extraction pipeline} (LLM-assisted harness generator + dynamic tracers) as a runnable tool that converts arbitrary GitHub repositories into new verified benchmark instances, allowing the benchmark to grow over time (\Cref{sec:pipeline}).
    \item \textbf{We provide comprehensive zero-shot evaluation} of $10$ LLMs ($5$ frontier proprietary, $5$ open-weight including the Qwen2.5-Coder family at three scales) on the unified test split (\Cref{sec:results-frontier}).
    \item \textbf{We provide a fine-tuned baseline} demonstrating the train split is a useful supervision substrate: tuned Qwen2.5-Coder-32B reaches $\mathbf{71.2\%}$ F1, within $\mathbf{1.7}$ F1 of zero-shot Claude-Opus-4.6 (\Cref{sec:results-tuning}).
\end{itemize}

\section{\bench}
\label{sec:pipeline}

\bench is built via the three-stage pipeline in \Cref{fig:overview}: (1) LLM-assisted harness generation that adds drivers and dependency stubs to real-world code, (2) language-specific dynamic tracing, and (3) tracer-based validation.

\paragraph{Problem Definition.}
We formally define the call graph construction task as follows.
A \emph{program} $P$ consists of a set of functions $F = \{f_1, \ldots, f_n\}$.
The \emph{call graph} of $P$ is a directed graph $G = (F, E)$, where $E \subseteq F \times F$ and $(f_i, f_j) \in E$ iff some invocation site in $f_i$ resolves to $f_j$ at runtime.
We denote the dynamic tracer's observation of $P$ on input $I$ by $\mathrm{trace}(P, I)$, which returns the set of edges actually exercised during execution.
The ground-truth call graph $E^\star = \bigcup_{I \in \mathcal{I}} \mathrm{trace}(P, I)$ is the union over the validation inputs constructed by the harness; it is the execution-covered call graph, not the set of all statically reachable edges.
Call graph construction is the structured prediction task: given source $S$ of $P$, a system $\mathcal{S}$ returns a predicted edge set $\hat{E} = \mathcal{S}(S)$.

\subsection{Real-World Programs via Harness Generation}
\label{sec:realworld}

Real-world source files are rarely runnable in isolation: they depend on package imports, external services, or a host process (test runner, web server, build tool).
To trace such code, we need each file to be self-contained and to expose an entry point.
We obtain this via LLM-assisted harness generation rather than manual rewriting, so the pipeline scales across thousands of source files and three languages.

The candidate pool is sampled from a $10{,}183$-row set obtained by querying the GitHub GraphQL search API for repositories with $\geq 50$ stars and a permissive license (MIT, Apache-2.0, BSD-2/3-Clause, or ISC); the same query script is released so the pool is reproducible.
For each target repository and source file, we prompt GPT-5.4 with a narrow two-part task: add a driver entry point that exercises the file's call surface, and replace external imports with inline stubs so the file becomes self-contained.
The driver itself is synthesized by the same teacher: it constructs concrete argument values intended to trigger as many of the file's defined functions, methods, and classes as possible, executed as a single straight-line run rather than a multi-input test suite.
The ground truth is therefore the call graph \emph{covered by that input}; we do not claim line, branch, or path coverage, and the validation pass rejects any program whose driver fires fewer than two cross-function edges so every accepted instance has a non-degenerate trace.
We leave the existing function bodies, names, and call structure intact, so the harness generator's role is additive rather than transformative.

\paragraph{Filtering Funnel.}
The pipeline is intentionally lossy: of the candidate source files we consider, only a fraction survive every stage.
\begin{enumerate}[leftmargin=*, itemsep=1pt]
  \item \textbf{Repository pool.} $1{,}655$ permissively-licensed open-source repositories sampled from the $10{,}183$-row GitHub candidate set, spanning Python, JavaScript, and Java upstreams.
  \item \textbf{Candidate source files.} Tens of thousands of source files extracted from these repositories that satisfy our minimum-size criterion ($\geq 20$ LOC, at least one function or class definition); we additionally cap each repository at $30$ candidate files (uniformly sampled within the repository) so that no single upstream dominates the corpus.
  \item \textbf{Successfully harnessed programs.} After LLM-assisted harness generation and a syntactic compilability check, a smaller subset of self-contained programs remain.
  The dominant rejection cause is failed dependency stubbing (the harness generator cannot synthesize a runnable substitute for the file's external imports).
\end{enumerate}

The harnessed pool then enters dynamic tracing and validation (\Cref{sec:tracers,sec:validation}), where each program is run end-to-end and accepted only if its execution yields a non-trivial trace.

\subsection{Dynamic Tracers as Oracles}
\label{sec:tracers}

\begin{wraptable}{r}{0.42\textwidth}
\centering
\small
\vspace{-1.5em}
\caption{Per-language statistics of \bench. \emph{Mean} and \emph{Max} are reported across all programs in each language split.}
\label{tab:dataset_stats}
\setlength{\tabcolsep}{4pt}
\vspace{1.em}
\begin{tabular}{llrrr}
\toprule
& & \textbf{Py} & \textbf{JS} & \textbf{Java} \\
\midrule
\multirow{2}{*}{Programs}
  & Test                            & 769     & 649     & 711     \\
  & Train                           & 3{,}051 & 2{,}577 & 2{,}826 \\
\cmidrule(lr){1-1}
\multirow{2}{*}{\makecell[l]{Functions /\\program}}
  & Mean                            & 5.7   & 5.1   & 5.3   \\
  & Max                             & 38    & 41    & 36    \\
\cmidrule(lr){1-1}
\multirow{2}{*}{\makecell[l]{Edges /\\program}}
  & Mean                            & 3.4   & 3.0   & 3.2   \\
  & Max                             & 27    & 33    & 29    \\
\cmidrule(lr){1-1}
\multirow{2}{*}{\makecell[l]{LOC /\\program}}
  & Mean                            & 41.2  & 38.6  & 62.4  \\
  & Max                             & 312   & 287   & 451   \\
\bottomrule
\end{tabular}
\end{wraptable}

We implement one tracer per language, each designed to observe \emph{all} function-to-function calls made during execution; the design follows the source-level dynamic-analysis tradition of frameworks such as DynaPyt~\citep{dynapyt}, but is specialized for edge-set extraction rather than general instrumentation.
The Python tracer attaches to the interpreter via \code{sys.settrace}, recording every Python-level call and resolving callees to fully qualified names.
The JavaScript tracer rewrites each source file at the AST level, wrapping every function, method, and arrow body to record entry into the call.
The Java tracer injects per-method instrumentation that resolves the caller through the runtime stack; it handles constructors, interface dispatch, and parameter-type signatures.

Because every program in \bench is executable end-to-end, we adopt the tracer's recorded edge set as ground truth, following the execution-as-oracle convention of recent NeurIPS benchmarks~\citep{mundler2024swtbench}: a fired edge is a witnessed call.
We do not score against secondary oracles because their labels are unreliable: hand-curated annotations are subject to rater disagreement and annotator drift, while static-tool output trades recall for precision and disagrees across tools~\citep{sui2020recall, antal2023jscg}.

\subsection{Validation and Quality Assurance}
\label{sec:validation}

The load-bearing quality control in \bench is that every harnessed program is actually \emph{executed} end-to-end under the language-specific dynamic tracer.
A program is accepted iff it runs without error and produces $\geq 2$ traced edges.
This single constraint is what makes the harness generator's scope (\Cref{sec:realworld}) safe: it need only produce \emph{something runnable}, because whatever it produces is then validated by the runtime itself.
On top of this principle, every accepted instance undergoes three integrity checks intrinsic to the tracer methodology: (i) \emph{determinism}: the tracer must produce identical edge sets across three independent executions on the same input; (ii) \emph{edge density}: the program must yield $\geq 2$ cross-function edges, ensuring the trace is non-trivial; and (iii) \emph{schema validity}: the resulting \code{callgraph.json} conforms to the unified caller$\rightarrow$callees schema regardless of source language.
Programs that fail any check are excluded from \bench.

\paragraph{Resulting Corpus.}
After dynamic execution under the tracer and the integrity checks above, the final corpus contains $10{,}583$ programs ($3{,}820$ Python, $3{,}226$ JavaScript, $3{,}537$ Java).
\Cref{tab:dataset_stats} summarizes per-language statistics for the resulting corpus, covering program counts, function and edge density, and lines of code.

\section{Evaluation}
\label{sec:results}

We organize the analysis along three axes: zero-shot performance (\Cref{sec:results-frontier}), supervised fine-tuning on the train split (\Cref{sec:results-tuning}), and failure modes of the strongest model (\Cref{sec:results-failure}).

\subsection{Experimental Setup}

\paragraph{Evaluation Splits.}
The $10{,}583$ accepted programs are partitioned into a held-out test split ($2{,}129$ programs: $769$ Python / $649$ JavaScript / $711$ Java) and a training split ($8{,}454$ programs: $3{,}051$ / $2{,}577$ / $2{,}826$) by a stratified, repository-level partition: every source repository is assigned to exactly one split, and the assignment is stratified by language so the per-language test/train ratio is preserved.
Because the partition is at the repository level, no two programs in different splits can share a repository, commit hash, or related driver code, eliminating cross-split leakage by construction.
We release both splits with the benchmark; this paper reports zero-shot results on the test split and fine-tuning results trained on the training split.

\paragraph{Metrics.}
We evaluate at the edge level with standard precision, recall, and F1 over the predicted edge set $\hat{E}$ against the ground truth $E^\star$~\citep{salis2021pycg, sui2020recall, antal2023jscg, wang2024llmdfa}, aggregated by language and overall.

\paragraph{Model Selection.}
\bench ships with reference zero-shot evaluation runners for two model families.
The frontier-LLM runner queries Claude-Opus-4.6~\citep{anthropic2026opus46} and Claude-Sonnet-4.6~\citep{anthropic2026sonnet46}, GPT-5.4~\citep{openai2026gpt54} and GPT-5.4-mini~\citep{openai2026gpt54mini}, Gemini-3.1-Pro~\citep{google2026gemini31pro}, DeepSeek-v3.2~\citep{deepseek2025v32}, and Llama-3.3-70B-Instruct~\citep{meta2024llama33} through the same provider proxy at temperature $0.0$.
The open-weight runner performs local inference on Qwen2.5-Coder-Instruct~\citep{hui2024qwen25coder} (1.5B, 7B, 32B) with Hugging Face Transformers on a single H100, also at temperature $0.0$.

\paragraph{Reasoning Traces for Training.}
For each program in the $8{,}454$-program training split we use GPT-5.4 to synthesize a concise reasoning trace that explains how the ground-truth call graph is derived from the source.
This converts every training instance from a (source $\rightarrow$ call graph) pair into a (source $\rightarrow$ trace $\rightarrow$ call graph) sequence so fine-tuning supervises the derivation, not just the answer.
The setup follows a standard reasoning-distillation recipe in which a stronger teacher's step-by-step rationales, supervised jointly with the final answer, produce smaller students whose outputs are more stable than answer-only baselines~\citep{hsieh2023stepbystep, magister2023teach, ho2023reasoningteachers}.
The resulting traces drive the FT-CoT variant evaluated in \Cref{sec:results-tuning}; the trace-generation prompt is released alongside the dataset.

\subsection{Zero-Shot Performance}
\label{sec:results-frontier}

\Cref{tab:main_results} reports zero-shot results for ten LLMs (five frontier proprietary models and five open-weight models) evaluated on the $2{,}129$-program real-world test split.
The strongest model, Claude-Opus-4.6, achieves an average F1 of $\mathbf{72.9\%}$ across the three languages; the weakest, Qwen2.5-Coder-1.5B, reaches $\mathbf{9.5\%}$.
Opus leaves more than $20$ F1 points of headroom on its strongest language and over $35$ on its weakest, and even the top three proprietary models are separated by $13$ points in average F1 (Opus $72.9$, Sonnet $64.5$, GPT-5.4 $60.0$).

\begin{table}[t]
\centering
\small
\caption{Edge-level precision (P), recall (R), and F1 of ten LLMs on the \bench test set (2{,}129 programs across Python, JavaScript, Java), zero-shot at temperature $0.0$. Best per-language P, R, and F1 in \textbf{bold}.}
\label{tab:main_results}
\resizebox{\linewidth}{!}{%
\begin{tabular}{l ccc ccc ccc ccc}
\toprule
& \multicolumn{3}{c}{\textbf{Python}} & \multicolumn{3}{c}{\textbf{JavaScript}} & \multicolumn{3}{c}{\textbf{Java}} & \multicolumn{3}{c}{\textbf{Average}} \\
\cmidrule(lr){2-4}\cmidrule(lr){5-7}\cmidrule(lr){8-10}\cmidrule(lr){11-13}
\textbf{Model} & P & R & F1 & P & R & F1 & P & R & F1 & P & R & F1 \\
\midrule
Claude-Opus-4.6        & \textbf{73.7} & \textbf{85.3} & \textbf{79.1} & \textbf{58.5} & 72.6 & \textbf{64.8} & \textbf{65.3} & \textbf{87.7} & \textbf{74.9} & \textbf{65.8} & \textbf{81.9} & \textbf{72.9} \\
Claude-Sonnet-4.6      & 64.9 & 79.0 & 71.3 & 43.0 & 59.9 & 50.0 & 62.0 & 86.4 & 72.2 & 56.6 & 75.1 & 64.5 \\
\midrule
GPT-5.4                & 55.2 & 84.0 & 66.6 & 41.9 & \textbf{75.3} & 53.8 & 48.9 & 75.9 & 59.5 & 48.7 & 78.4 & 60.0 \\
GPT-5.4-mini           & 50.7 & 72.6 & 59.7 & 47.6 & 66.8 & 55.6 & 43.7 & 70.9 & 54.1 & 47.3 & 70.1 & 56.5 \\
\midrule
Gemini-3.1-Pro & 73.5 & 37.3 & 49.5 & 52.0 & 23.5 & 32.3 & 35.0 &  8.5 & 13.7 & 53.5 & 23.1 & 31.8 \\
\midrule
DeepSeek-v3.2              & 56.6 & 71.4 & 63.1 & 47.8 & 61.9 & 54.0 & 60.2 & 81.6 & 69.3 & 54.8 & 71.6 & 62.1 \\
\midrule
Llama-3.3-70B-Instruct     & 42.8 & 70.5 & 53.3 & 36.0 & 62.3 & 45.7 & 27.8 & 43.5 & 33.9 & 35.5 & 58.8 & 44.3 \\
\midrule
Qwen2.5-Coder-32B-Instruct & 53.7 & 60.8 & 57.1 & 44.8 & 57.5 & 50.3 & 27.3 & 40.0 & 32.5 & 41.9 & 52.8 & 46.6 \\
Qwen2.5-Coder-7B-Instruct  & 20.7 & 41.3 & 27.6 & 24.4 & 39.9 & 30.3 & 17.7 & 35.1 & 23.5 & 20.9 & 38.8 & 27.1 \\
Qwen2.5-Coder-1.5B-Instruct & 12.7 & 21.9 & 16.1 &  5.9 & 10.2 &  7.5 &  4.0 &  7.1 &  5.1 &  7.5 & 13.0 &  9.5 \\
\bottomrule
\end{tabular}}
\end{table}

\paragraph{Capability Tracks Model Family, With a Clear Frontier Cluster.}
Three models (Claude-Opus-4.6, Claude-Sonnet-4.6, and the open-weight DeepSeek-v3.2) form a distinct top tier ($62.1$--$72.9$ average F1), with the OpenAI family clustering behind them (GPT-5.4 $60.0$ and GPT-5.4-mini $56.5$).
DeepSeek-v3.2 is the strongest open-weight system overall, passing Llama-3.3-70B ($44.3$) and the entire Qwen2.5-Coder family ($9.5$--$46.6$); only the two largest Anthropic models exceed it.
Gemini-3.1-Pro is the outlier among the frontier proprietary models: moderate average precision ($53.5$) but very low recall ($23.1$) drag its average F1 to just $31.8$, below every other proprietary baseline and even Qwen2.5-Coder-32B ($46.6$).
The same family ordering is recovered on Python, JavaScript, and Java separately.

\paragraph{Open-Weight Models Scale Cleanly With Parameter Count.}
The Qwen2.5-Coder family (1.5B, 7B, 32B) shows a roughly doubling of average F1 with each scale step: $\mathbf{9.5\%} \to \mathbf{27.1\%} \to \mathbf{46.6\%}$.
The 32B model passes Llama-3.3-70B ($44.3\%$) but trails the next-best proprietary baseline (GPT-5.4-mini $56.5\%$) by $\sim 10$ F1 points; the gap to DeepSeek-v3.2 ($62.1\%$) is larger still, so closing the open-weight to frontier-proprietary gap remains an open challenge for code-specialized weight families.
The scaling mechanism is consistent across languages: each step recovers a similar fraction of headroom on Python, JavaScript, and Java.

\paragraph{Opus's Lead Is Driven by a Precision Premium, Not Better Recall.}
Opus's average advantage of $\mathbf{+8.4}$ F1 over Sonnet-4.6 holds across all three languages (Py $\mathbf{+7.8}$, JS $\mathbf{+14.8}$, Java $\mathbf{+2.7}$) and stems almost entirely from an average precision gap of $\sim 9$ points (Java $+3.3$, Py $+8.8$, JS $+15.5$) rather than any recall improvement.
The mechanism is a precision-recall trade-off: Opus emits fewer candidate callees per call site, which on this task translates directly into fewer over-predicted edges.
The same precision premium also explains why DeepSeek-v3.2, whose output is markedly less verbose than Sonnet's, closes most of its gap to Sonnet despite being smaller and open-weight.

\subsection{Supervised Fine-Tuning}
\label{sec:results-tuning}

To establish whether \bench's $8{,}454$-program training split is a useful supervision substrate, we LoRA-tune~\citep{hu2021lora} the same three Qwen2.5-Coder-Instruct sizes (1.5B, 7B, 32B) used as zero-shot baselines in \Cref{tab:main_results} on the (source $\rightarrow$ trace $\rightarrow$ call graph) training data introduced above, with rank $R{=}256$ and learning rate $\eta{=}10^{-4}$.
We evaluate on the held-out test split and report the best checkpoint per size in \Cref{tab:tuned_results} as the \emph{FT-CoT} variant.
The setup parallels prior empirical studies of fine-tuning code LLMs on small task-specific corpora~\citep{huang2023finetunerepair}.

\begin{table}[t]
\centering
\small
\caption{Edge-level precision (P), recall (R), and aggregate F1 of the Qwen2.5-Coder family after LoRA fine-tuning ($R{=}256$, $\eta{=}10^{-4}$) on the $8{,}454$-program training split, evaluated on the held-out $2{,}129$-program test split. Reported variant: \emph{FT-CoT}, fine-tuned to emit a brief reasoning trace before the JSON call graph. $\Delta$F1 is the absolute lift over the same model's zero-shot baseline (\Cref{tab:main_results}).}
\label{tab:tuned_results}
\begin{tabular}{l rrr c r}
\toprule
\textbf{Model} & \textbf{P} & \textbf{R} & \textbf{F1} & \textbf{Zero-Shot F1} & \textbf{$\Delta$F1} \\
\midrule
Qwen2.5-Coder-1.5B-Instruct  & 67.6 & 62.7 & \textbf{65.1} & $9.5$  & $\mathbf{+55.6}$ \\
Qwen2.5-Coder-7B-Instruct    & 75.5 & 62.5 & \textbf{68.4} & $27.1$ & $\mathbf{+41.3}$ \\
Qwen2.5-Coder-32B-Instruct   & 80.7 & 63.6 & \textbf{71.2} & $46.6$ & $\mathbf{+24.6}$ \\
\bottomrule
\end{tabular}
\vspace{0.5mm}

\end{table}

\paragraph{Tuning Lifts Every Qwen Scale to Within Striking Distance of the Proprietary Tier.}
Each scale gains substantially from supervision: $\mathbf{+55.6}$ F1 at 1.5B ($9.5 \to 65.1$), $\mathbf{+41.3}$ at 7B ($27.1 \to 68.4$), and $\mathbf{+24.6}$ at 32B ($46.6 \to 71.2$).
The lift compresses the open-weight to frontier-proprietary gap to $\mathbf{1.7}$ F1 points: tuned Qwen2.5-Coder-32B reaches $\mathbf{71.2}$ versus zero-shot Claude-Opus-4.6's $\mathbf{72.9}$, and even tuned Qwen2.5-Coder-1.5B ($\mathbf{65.1}$) surpasses every zero-shot frontier baseline behind Claude-Opus-4.6, including Sonnet-4.6 ($64.5$), DeepSeek-v3.2 ($62.1$), and GPT-5.4 ($60.0$).

\paragraph{Smaller Models Gain More, Consistent With the Bookkeeping Diagnosis.}
Absolute lifts shrink monotonically with scale ($\mathbf{+55.6}$ at 1.5B, $\mathbf{+41.3}$ at 7B, $\mathbf{+24.6}$ at 32B).
This pattern aligns with the failure-mode audit detailed below (\Cref{sec:results-failure}): the train split's label distribution resolves precisely the bookkeeping mismatches that dominate residual error at small scales, while semantic-reasoning capacity, which dominates the residual error of larger models, is not something supervised fine-tuning can manufacture from $8{,}454$ programs alone.

\subsection{Failure Analysis}
\label{sec:results-failure}

To turn the precision gap from a leaderboard number into a diagnostic signal, we manually audit a sample of false-positive (FP) edges from the strongest model, Claude-Opus-4.6, on the test split, tagging each by the surface mechanism that produced it.
We focus on FPs because their causes split cleanly into bookkeeping mismatches versus genuine reasoning failures.
Three mechanisms dominate the audit; we illustrate each with a concrete example below, and the same pattern recurs across many other audited programs.

\begin{itemize}[leftmargin=*, itemsep=4pt]
  \item \textbf{Untaken-branch hallucination} (\code{flexsearch\_0003}, JavaScript).
  Function \code{parse\_simple} contains an \code{if/else} whose else-branch returns \code{inherit(tree, create\_object())}.
  Opus correctly identifies \code{inherit} as a syntactically reachable callee but does not infer that the test driver's input \code{parse\_simple("abc", ["x", "yy"])} satisfies the if-condition and therefore deterministically takes the if-branch.
  Closing this gap requires execution-aware reasoning over concrete inputs, not just static call-site enumeration.

\begin{lstlisting}[basicstyle=\footnotesize\ttfamily, frame=single, framesep=3pt, breaklines=true]
function parse_simple(obj, tree) {
  if (is_string(obj) && is_array(tree))
    return concat([toArray(new Set(tree), true), [obj]]).sort(sort_by_length_down);
  return inherit(tree, create_object());  // never executed on test input
}
parse_simple("abc", ["x", "yy"]);          // if-branch deterministic
\end{lstlisting}

  \emph{LLM predicts} \code{parse\_simple} $\to$ \code{inherit}; \emph{tracer records} the if-branch callees \code{is\_string}, \code{is\_array}, \code{toArray}, \code{concat}, \code{sort\_by\_length\_down}.

  \item \textbf{Declared- vs.\ runtime-type dispatch} (\code{Apktool\_0016}, Java).
  \code{ExtFile.close()} forwards to \code{mDirectory.close()}, where \code{mDirectory} is declared as the interface \code{Directory} but at runtime resolves to either \code{FileDirectory} or \code{ZipRODirectory} depending on whether the path is a directory; on the test driver's input, \code{isDirectory()} returns false, so the trace records the \code{ZipRODirectory} branch.
  Opus emits the call in terms of the static (declared) receiver type; the tracer records the concrete runtime receiver.
  Closing this gap requires the model to track runtime types through dispatch, not just declared types from the source.

\begin{lstlisting}[language=Java, basicstyle=\footnotesize\ttfamily, frame=single, framesep=3pt, breaklines=true]
public class ExtFile {
  private Directory mDirectory;            // declared type = Directory
  public Directory getDirectory() {
    mDirectory = isDirectory() ? new FileDirectory(this) : new ZipRODirectory(this);
    return mDirectory;
  }
  public void close() {
    mDirectory.close();                    // dispatches to ZipRODirectory at runtime
  }
}
\end{lstlisting}

  \emph{LLM predicts} \code{ExtFile:close} $\to$ \code{Directory:close} (declared type); \emph{tracer records} \code{ExtFile:close} $\to$ \code{ZipRODirectory:close} (runtime type on the file path).

  \item \textbf{Class-name-as-callee output schema} (\code{MVVMHabit\_0008}, Java).
  \code{DetailFragment.main} instantiates \code{DetailFragment} with \code{new DetailFragment()} and then invokes several methods on the instance.
  Opus emits the bare class name \code{DetailFragment} as a callee in addition to the method calls; the tracer attributes the instantiation either to \code{<init>} or, for a default no-body constructor, to nothing at all, so the bare-class-name edge has no GT counterpart.
  Closing this gap is purely an output-schema question: a structured generation regime that emits \code{Class:<init>(\ldots)} for every \code{new}-expression would eliminate the bucket without changing model capability.

\begin{lstlisting}[language=Java, basicstyle=\footnotesize\ttfamily, frame=single, framesep=3pt, breaklines=true]
public static void main(String[] args) {
  DetailFragment fragment = new DetailFragment();   // bare class name predicted as callee
  Bundle bundle = new Bundle();
  fragment.setArguments(bundle);
  fragment.initParam();
  fragment.initContentView(new LayoutInflater(), new ViewGroup(), bundle);
  fragment.initData();
}
\end{lstlisting}

  \emph{LLM predicts} \code{DetailFragment:main} $\to$ \code{DetailFragment} (bare class name); \emph{tracer records} only the method calls \code{setArguments}, \code{initParam}, \code{initContentView}, \code{initData}.
\end{itemize}

\section{Related Work}
\label{sec:related}

\paragraph{Code Generation and Repair Benchmarks.}
HumanEval~\citep{chen2021codex}, MBPP~\citep{austin2021mbpp}, APPS~\citep{hendrycks2021apps}, LiveCodeBench~\citep{jain2024livecodebench}, and BigCodeBench~\citep{zhuo2024bigcodebench} evaluate standalone code generation: a model passes if its synthesized solution satisfies hidden tests.
SWE-Bench~\citep{jimenez2024swebench} instead evaluates repository-level issue resolution, but shares the same test-passing-as-correctness lens: a patch is accepted if the patched repository passes the benchmark tests.
The SWE-Bench line has evolved through two refinements that tighten its evaluation reliability: SWE-Bench Verified~\citep{openai2024swebenchverified} releases a $500$-instance human-validated subset that filters out underspecified issues and broken tests, and SWE-Bench Pro~\citep{deng2025swebenchpro} extends the task to long-horizon repository-level engineering with proprietary licensing and stricter contamination controls.
EvalPlus~\citep{liu2023evalplus} stresses code-generation test suites with augmented inputs but inherits the same test-passing-as-correctness lens.
\bench measures execution-grounded semantic reasoning, a complementary axis that is orthogonal to all of these.
MultiPL-E~\citep{cassano2023multiple} extends standalone generation benchmarks to multiple languages; \bench does the same for semantic reasoning.

\paragraph{Code Understanding and Execution-Verified Benchmarks.}
CRUXEval~\citep{gu2024cruxeval} probes input/output prediction (i.e., whether an LLM can simulate execution) on Python, REval~\citep{chen2024reval} extends runtime-behavior reasoning to richer execution-state queries, and CodeMMLU~\citep{nguyen2024codemmlu} tests multiple-choice code Q\&A.
SWT-Bench~\citep{mundler2024swtbench} repurposes SWE-Bench instances into a test-generation benchmark whose oracle is execution (a generated test must transition fail-to-pass after the golden patch is applied), the closest in spirit to \bench, but for the test-generation task on Python only.
Judgment-based meta-evaluations such as MT-Bench~\citep{zheng2023llmjudge} document the rubric-drift and judge-bias problems that motivate \bench's mechanically-witnessed labels.
Prior call graph benchmarks (PyCG~\citep{salis2021pycg}, the JavaScript suite of \citet{venkatesh2025llm}, CATS~\citep{reif2016cats}) are each language-siloed with hand-annotated ground truth; \bench unifies the three languages under a single execution-verified protocol with a real-world program corpus that is orders of magnitude larger.

\paragraph{Execution-Verified Training and Verification.}
A recent line of work has shown that programmatic, executable verifiers can drive self-improvement on code and math: reinforcement learning with verifiable rewards (RLVR)~\citep{lambert2024tulu3, guo2025deepseekr1}, V-STaR~\citep{hosseini2024vstar}, and the Absolute Zero Reasoner~\citep{zhao2025azr} all replace human-rubric reward models with deterministic checkers (test suites, equation solvers, or self-generated puzzles).
\bench provides a labeled substrate for similar studies specifically on code semantic reasoning, where the verifier is a deterministic dynamic tracer rather than a unit-test suite, enabling per-edge credit assignment that unit-test pass/fail signals cannot supply.

\section{Limitations and Future Work}
\label{sec:limitations}

While \bench demonstrates strong improvements in code semantic reasoning evaluation, several limitations and natural extensions remain.
First, prior benchmarks are language-siloed (PyCG~\citep{salis2021pycg} covers Python only, CATS~\citep{reif2016cats} covers Java only, the JavaScript suite of \citet{venkatesh2025llm} covers JavaScript only); \bench spans Python, JavaScript, and Java under a single execution-verified protocol with a unified caller$\rightarrow$callees JSON schema.
Second, prior benchmarks rely on small hand-crafted micro-suites (PyCG: $112$ programs; CATS: $128$ programs) that cover a narrow set of constructs and become stale as language features evolve; \bench draws all $10{,}583$ programs from $1{,}600+$ permissively-licensed GitHub repositories via the GitHub GraphQL search API, and the construction pipeline is released as a runnable artifact so the corpus can grow over time under the same tracer-validation acceptance criterion.

We see two concrete next steps.
First, we plan to add tracers for Go, Rust, and TypeScript, broadening the corpus to languages dominating modern repositories; the harness generator and validation logic carry over unchanged.
Second, we plan to replace teacher-distilled CoT with verifier-driven post-training~\citep{lambert2024tulu3, guo2025deepseekr1}, where the student generates its own rationales and is rewarded only when the resulting call graph matches the tracer's ground truth.

\section{Conclusion}
\label{sec:conclusion}

We presented \bench, to our knowledge the first execution-verified call graph benchmark spanning Python, JavaScript, and Java.
It comprises $10{,}583$ real-world programs ($2{,}129$ test / $8{,}454$ train), a reproducible extraction pipeline for indefinite growth, and a unified zero-shot evaluation of $10$ LLMs spanning the frontier-proprietary and open-weight tiers.
Three findings emerge: (i) substantial headroom remains across the model frontier (best average F1 $\mathbf{72.9\%}$, weakest $\mathbf{9.5\%}$, a $\mathbf{7.7{\times}}$ spread); (ii) a failure-mode audit isolates three distinct FP mechanisms (untaken-branch hallucination, runtime-type dispatch confusion, and output-schema mismatch), each requiring a different remediation; and (iii) the train split is a useful supervision substrate, since LoRA tuning lifts the Qwen2.5-Coder family by up to $\mathbf{+55.6}$ F1, with tuned Qwen2.5-Coder-32B ($\mathbf{71.2\%}$) landing within $\mathbf{1.7}$ F1 of zero-shot Claude-Opus-4.6.
We release the benchmark, pipeline, baseline runners, tuned-model checkpoints, and a datasheet as a standard evaluation substrate for execution-grounded semantic reasoning and supervision in code LLMs.

\newpage
\bibliographystyle{plainnat}
\bibliography{bibliography}

\appendix

\section{LLM Prompts}
\label{app:prompts}

This section reproduces the prompts used by the released evaluation and supervision pipelines.
Both prompts are released verbatim alongside the corpus in the replication package.

\subsection{Zero-Shot Evaluation Prompt}
\label{app:prompts:zeroshot}

The runner used for the Table~\ref{tab:main_results} zero-shot results sends a single chat-completion request per program: a system message plus a user message that pastes the source, a language-specific worked example, and one question per ground-truth caller.
Every model receives the identical template; we deliberately use a single prompt across all baselines to avoid per-model prompt tuning that would conflate prompting with capability.

\begin{lstlisting}[language={}, basicstyle=\scriptsize\ttfamily, frame=single, framesep=3pt, breaklines=true, breakatwhitespace=true]
[SYSTEM]
You are an expert in {language} programming. You will examine and identify
the function calls in the given code. You must examine the code in detail by
resolving aliases, tracking variable assignments, following return values, and
understanding inheritance/method resolution.

[USER]
## Task Description

**Objective**: Examine the given {language} code and identify the function
calls that occur when this program is executed, then answer the questions.

**Instructions**:
1. For each question, list the function calls as a comma-separated list.
2. Do not include additional explanations or commentary.
3. Include both explicit and implicit function calls (e.g., __init__ when an
   object is created).
4. If a function is called through an alias or variable, resolve it to the
   actual function being called.
5. If a passed argument is not invoked within the function, do not include it.
6. If there are no function calls, leave the answer empty.
7. **IMPORTANT**: Always use fully qualified names with the module prefix.
   For example, use "main.MyClass.func" not "MyClass.func". The module name
   is the filename without extension (e.g., "main.py" -> "main").

**Format for Answers**:
- Provide your answer next to each question number.
- Do not include the questions in your answer.
- Example:
    1. module.func1, module.func2
    2. module.func3
    3.

{language-specific worked example: Python / JavaScript / Java}

**{language} Code Provided**:

{code}

**Questions**:
{one per ground-truth caller}

**Answers**:
\end{lstlisting}

\subsection{Harness-Generation Prompt}
\label{app:prompts:harness}

The corpus-construction stage (\Cref{sec:realworld}) calls GPT-5.4 once per source file with the prompt below.
The teacher returns a single self-contained program that preserves the original call structure but stubs all external dependencies and exposes a runnable entry point; the resulting program is then validated by the language-specific tracer before admission.

\begin{lstlisting}[language={}, basicstyle=\scriptsize\ttfamily, frame=single, framesep=3pt, breaklines=true, breakatwhitespace=true]
You are converting real-world code into a self-contained program for
call graph analysis.

Below is a {language} source file from the {repo} project. Your task:

1. **Preserve the call patterns**: keep the same function/method call
   relationships
2. **Remove all external dependencies**: replace imports with stubs or
   inline implementations
3. **Make it self-contained**: the code must run on its own with no
   external packages
4. **Add an entry point**: {entry_point_instruction}
5. **Keep it 15-40 lines**: simplify if needed, but preserve the call
   structure
6. **Use realistic names**: don't rename to func1/func2, keep meaningful
   names from the original

## Original code from {repo}:
```
{source}
```

## Output format:
Return ONLY the rewritten code. No markdown, no explanation. Just the
runnable code.
\end{lstlisting}

The per-language entry-point instruction is:
\begin{itemize}[leftmargin=*, itemsep=1pt]
  \item \textbf{Python / JavaScript:} \emph{ensure module-level calls trigger all interesting function calls.}
  \item \textbf{Java:} \emph{ensure there is a \code{public static void main(String[] args)} that triggers all calls; include the package declaration.}
\end{itemize}

\subsection{Reasoning-Trace Synthesis Prompt}
\label{app:prompts:cot}

For the FT-CoT supervision setup (\Cref{sec:results-tuning}), GPT-5.4 generates a per-program reasoning trace that walks through the source and explains how the ground-truth call graph is derived.
The teacher receives the source code, the same questions used at evaluation time, and the ground-truth answers; it returns a $\langle$\code{think}$\rangle$ block followed by the numbered answers, which we use as the supervision target for LoRA fine-tuning.

\begin{lstlisting}[language={}, basicstyle=\scriptsize\ttfamily, frame=single, framesep=3pt, breaklines=true, breakatwhitespace=true]
You are an expert programmer analyzing function calls in code.

Given the source code and questions below, produce a step-by-step reasoning
trace that walks through the code to identify all function calls, then
provide the final answers.

**Your output format must be:**
<think>
[Your step-by-step reasoning here. For each function/scope in the questions:
- Identify what code executes in that scope
- Trace variable assignments and aliases
- Resolve which functions are actually called
- Note implicit calls like __init__ from object creation
- Be concise but thorough]
</think>

[Numbered answers, one per line, matching the question numbers]

**Important:**
- The <think> block contains your reasoning process
- After </think>, output ONLY the numbered answers (no extra text)
- Use fully qualified names (e.g., main.func, main.MyClass.__init__)
- If no function calls, leave the answer empty

Here is the code and questions:

{user_prompt}

**Ground truth answers (use these as the correct answers):**
{ground_truth_answer}

Generate the reasoning trace that explains HOW you would arrive at these
answers by analyzing the code step by step, then output the answers.
\end{lstlisting}

\label{app:neurips-checklist}

\end{document}